\documentclass[aps,prl,twocolumn,showpacs,lengthcheck,preprintnumbers]{revtex4}%
\usepackage{amsfonts}
\usepackage{amsmath}
\usepackage{amssymb}
\usepackage{graphicx}%
\begin{document}
\preprint{Phys. Rev. A \textbf{77}, 013617 (2008)}
\title{Effective mean-field equations for cigar-shaped and disk-shaped Bose-Einstein condensates}
\author{A. Mu\~{n}oz Mateo}
\email{ammateo@ull.es}
\author{V. Delgado}
\email{vdelgado@ull.es}
\affiliation{Departamento de F\'{\i}sica Fundamental II, Universidad de La Laguna, La
Laguna, Tenerife, Spain}
\date{17 July 2007}


\begin{abstract}
By applying the standard adiabatic approximation and using the accurate
analytical expression for the corresponding local chemical potential obtained
in our previous work [Phys. Rev. A \textbf{75}, 063610 (2007)] we derive an
effective 1D equation that governs the axial dynamics of mean-field
cigar-shaped condensates with repulsive interatomic interactions, accounting
accurately for the contribution from the transverse degrees of freedom. This
equation, which is more simple than previous proposals, is also more accurate.
Moreover, it allows treating condensates containing an axisymmetric vortex
with no additional cost. Our effective equation also has the correct limit in
both the quasi-1D mean-field regime and the Thomas-Fermi regime and permits
one to derive fully analytical expressions for ground-state properties such as
the chemical potential, axial length, axial density profile, and local sound
velocity. These analytical expressions remain valid and accurate in between
the above two extreme regimes. Following the same procedure we also derive an
effective 2D equation that governs the transverse dynamics of mean-field
disk-shaped condensates. This equation, which also has the correct limit in
both the quasi-2D and the Thomas-Fermi regime, is again more simple and
accurate than previous proposals. We have checked the validity of our
equations by numerically solving the full 3D Gross-Pitaevskii equation.

\end{abstract}
\maketitle



\section{I. INTRODUCTION}

The experimental realization of Bose-Einstein condensates (BECs) of dilute
atomic gases confined in optical and magnetic traps \cite{BEC1,BEC2,BEC3} has
opened new opportunities for investigating the coherence properties of
degenerate quantum systems. From a theoretical point of view, under the usual
experimental conditions these systems can be accurately described by the
Gross-Pitaevskii equation (GPE) \cite{GPE}, a mean-field equation of motion
governing the behavior of the condensate wave function $\psi(\mathbf{r},t)$%
\begin{equation}
i\hbar\frac{\partial\psi}{\partial t}=\left(  -\frac{\hbar^{2}}{2m}\nabla
^{2}+V(\mathbf{r})+gN\left\vert \psi\right\vert ^{2}\right)  \psi. \label{I-1}%
\end{equation}
In the above equation $N$ is the number of atoms, $g=4\pi\hbar^{2}a/m$ is the
interaction strength, $a$ is the \textsl{s}-wave scattering length, and
$V(\mathbf{r})$ is the potential of the confining trap. In what follows we
shall restrict ourselves to the usual case of condensates with repulsive
interatomic interactions ($a>0$).

The GPE has proved to be very successful in describing the evolution in time
of dilute quantum gases near the zero-temperature limit. From a mathematical
point of view this equation is a time-dependent nonlinear differential
equation. Since no explicit analytical solutions are known, in general, Eq.
(\ref{I-1}) has to be solved numerically. This is a nontrivial numerical task
that demands a considerable computational effort. Moreover, in many
circumstances the superfluid dynamics of a zero-temperature condensate can
become chaotic which requires large basis or grid point sets to guarantee
convergence \cite{PRL06}. In recent years there has been particular interest
in BECs confined in highly anisotropic traps
\cite{Olsha1,Petrov1,Petrov2,Dunj1,Das1,Kett1,Strin1}. In such geometries the
condensate is so tightly confined in the radial or the axial dimension that
the corresponding dynamics becomes effectively one dimensional or two
dimensional, respectively. The time evolution of these systems with reduced
dimensionality is characterized by two very different time scales. Even though
usually one is only interested in the evolution of the slow degrees of freedom
in the effective mean field induced by the fast degrees of freedom, from a
computational point of view it is also required to resolve accurately the
irrelevant fast degrees of freedom. It is clear that for sufficiently
anisotropic traps this can represent a computational challenge. It is
therefore convenient to develop theoretical models that permit one to study
the condensate dynamics in terms of effective equations of lower
dimensionality. In this regard various approaches have been followed in recent
years \cite{Jack1,Chio1,Reatto1,Modug1,Kam1,You1}. Among them,\ the effective
1D and 2D nonpolynomial nonlinear Schr\"{o}dinger equations by Salasnich
\emph{et al. }\cite{Reatto1} have proved the most efficient.

In this work we derive effective 1D and 2D wave equations that govern the
dynamics of mean-field cigar-shaped and disk-shaped condensates with repulsive
interatomic interactions. These equations which incorporate properly the
contribution from the fast degrees of freedom have the correct limits in both
the TF and the perturbative regime. Even though the equations by Salasnich
\emph{et al. }are in general very accurate, we demonstrate that our effective
equations are more accurate. Moreover, as a consequence of its simplicity,
they also permit one to obtain fully analytical expressions for various
relevant ground-state properties such as chemical potentials, condensate
lengths, density profiles, and local sound velocities.

\section{II. CIGAR-SHAPED CONDENSATES}

Consider a BEC confined in a highly elongated trap. The high anisotropy of the
trap has important consequences on the condensate dynamics which under usual
conditions becomes governed by two very different time scales. In these
circumstances the characteristic evolution time of the fast transverse motion
($\sim\omega_{\bot}^{-1}$) is so small in comparison with the characteristic
time scale of the axial motion ($\sim\omega_{z}^{-1}$) that one can assume
that at every instant of time the transverse degrees of freedom adjust
instantaneously to the lowest-energy configuration compatible with the axial
configuration occurring at that time (\emph{adiabatic approximation}). This
implies, in particular, that the correlations between transverse and axial
motions can be neglected and the condensate wave function can be factorized as
\cite{Jack1,Kramer1}%
\begin{equation}
\psi(\mathbf{r},t)=\varphi(\mathbf{r}_{\bot};n_{1}(z,t))\phi(z,t),
\label{II-1}%
\end{equation}
where $\mathbf{r}_{\bot}=(x,y)$ and $n_{1}(z,t)$ is the local condensate
density per unit length characterizing the axial configuration%
\begin{equation}
n_{1}(z,t)\equiv N\int d^{2}\mathbf{r}_{\bot}|\psi(\mathbf{r}_{\bot}%
,z,t)|^{2}. \label{II-2}%
\end{equation}
Normalizing the transverse wave function to unity%
\begin{equation}
\int d^{2}\mathbf{r}_{\bot}|\varphi(\mathbf{r}_{\bot};n_{1})|^{2}=1,
\label{II-3}%
\end{equation}
Eq. (\ref{II-2}) takes the desirable form%
\begin{equation}
n_{1}(z,t)=N|\phi(z,t)|^{2}. \label{II-4}%
\end{equation}

Substituting now the wave function (\ref{II-1}) into the GPE and assuming the
confining potential to be separable as $V(\mathbf{r})=V_{\bot}(\mathbf{r}%
_{\bot})+V_{z}(z)$, one obtains%
\begin{multline}
\left(  i\hbar\frac{\partial\phi}{\partial t}+\frac{\hbar^{2}}{2m}%
\frac{\partial^{2}\phi}{\partial z^{2}}-V_{z}(z)\phi\right)  \varphi
(\mathbf{r}_{\bot};n_{1})=\\
\left(  -\frac{\hbar^{2}}{2m}\nabla_{\bot}^{2}\varphi+V_{\bot}(\mathbf{r}%
_{\bot})\varphi+gn_{1}(z,t)\left\vert \varphi\right\vert ^{2}\varphi\right)
\phi(z,t).\label{II-5}%
\end{multline}
Both the axial and time variations induced in the transverse wave function
$\varphi$ by the axial density $n_{1}$ have been neglected in the above
equation. Neglecting the time derivative of $\varphi$ is in fact the essence
of the adiabatic approximation already discussed. Neglecting the second
derivative of $\varphi$ with respect to $z$ requires the axial density to vary
sufficiently slowly along the axial direction. For condensates in highly
elongated traps such condition usually holds in most cases of practical interest.

Multiplying Eq. (\ref{II-5}) by $\varphi^{\ast}(\mathbf{r}_{\bot};n_{1})$ and
integrating on the transverse coordinates $\mathbf{r}_{\bot}$ we arrive at%
\begin{equation}
i\hbar\frac{\partial\phi}{\partial t}=-\frac{\hbar^{2}}{2m}\frac{\partial
^{2}\phi}{\partial z^{2}}+V_{z}(z)\phi+\mu_{\bot}(n_{1})\phi,\label{II-6}%
\end{equation}
where we have defined%
\begin{equation}
\mu_{\bot}(n_{1})\!\equiv\!\!\int\!d^{2}\mathbf{r}_{\bot}\varphi^{\ast
}\!\left(  \!-\frac{\hbar^{2}}{2m}\nabla_{\bot}^{2}+V_{\bot}(\mathbf{r}_{\bot
})+gn_{1}\left\vert \varphi\right\vert ^{2}\!\right)  \!\varphi.\label{II-7}%
\end{equation}
Equation (\ref{II-6}) is an effective 1D mean-field equation that governs the
axial dynamics of cigar-shaped BECs, incorporating the contribution from the
transverse degrees of freedom through $\mu_{\bot}(n_{1})$. Substitution of Eq.
(\ref{II-6}) into Eq. (\ref{II-5}) yields%
\begin{equation}
\left(  -\frac{\hbar^{2}}{2m}\nabla_{\bot}^{2}+V_{\bot}(\mathbf{r}_{\bot
})+gn_{1}\left\vert \varphi\right\vert ^{2}\right)  \varphi=\mu_{\bot}%
(n_{1})\varphi,\label{II-8}%
\end{equation}
which shows that at every instant of time and $z$ plane the transverse wave
function $\varphi$ satisfies the \emph{stationary} GPE of an axially
homogeneous condensate characterized by a density per unit length $n_{1}%
(z,t)$, with $\mu_{\bot}(n_{1})$ being the corresponding (transverse) local
chemical potential. Clearly, the usefulness of the effective 1D axial equation
(\ref{II-6}) depends, to a great extent, on the possibility of finding a
simple way of solving Eq. (\ref{II-8}). In this regard, it is important to
note that $n_{1}$ enters the equation above as a mere external parameter, so
that, in practice, the solution of the transverse equation does not require
the knowledge of the axial evolution. Before considering this problem in more
detail we will first take a look at the condensate stationary states, which
must satisfy%
\begin{equation}
\phi(z,t)=\phi_{0}(z)e^{-i\mu t/\hbar}.\label{II-9}%
\end{equation}
Substituting in Eq. (\ref{II-6}) one obtains%
\begin{equation}
-\frac{\hbar^{2}}{2m}\frac{\partial^{2}\phi_{0}}{\partial z^{2}}+V_{z}%
(z)\phi_{0}+\mu_{\bot}(n_{1})\phi_{0}=\mu\phi_{0},\label{II-10}%
\end{equation}
where the condensate chemical potential$\ \mu$ is the Lagrange parameter that
guarantees the normalization condition $\int dz|\phi(z,t)|^{2}=1$. When
$\phi_{0}$ varies so slowly that the typical length scale $\Delta_{z}$ of its
spatial variations along $z$ is much greater than the corresponding
\emph{axial healing length}, i.e.,%
\begin{equation}
\Delta_{z}\gg\frac{\hbar}{\sqrt{2m(\mu_{\bot}(n_{1})-\hbar\omega_{\bot})}%
},\label{II-11}%
\end{equation}
then the first term on the left in Eq. (\ref{II-10}) can be neglected in
comparison with the mean-field interaction energy and one arrives at the local
density approximation%
\begin{equation}
\mu=\mu_{\bot}(n_{1})+V_{z}(z).\label{II-12}%
\end{equation}
When this equation is applicable, the knowledge of the transverse chemical
potential permits one to obtain an analytic expression for the ground-state
axial condensate profile $n_{1}(z)$. In any case, as Eq. (\ref{II-6}) shows,
the analytic determination of $\mu_{\bot}(n_{1})$ is the key ingredient to
derive a useful 1D effective equation of motion for the condensate axial dynamics.

We shall concentrate in what follows on cigar-shaped condensates confined in
the radial direction by an axisymmetric harmonic potential characterized by an
oscillator length $a_{\bot}=\sqrt{\hbar/m\omega_{\bot}}$,%
\begin{equation}
V(\mathbf{r})=\frac{1}{2}m\omega_{\bot}^{2}r_{\bot}^{2}+V_{z}(z).\label{II-13}%
\end{equation}
Introducing dimensionless variables $\overline{r}_{\bot}=r_{\bot}/a_{\bot}$
and $\overline{\varphi}=a_{\bot}\varphi$, the transverse equation (\ref{II-8})
determining the local chemical potential $\overline{\mu}_{\bot}=\mu_{\bot
}/\hbar\omega_{\bot}$ takes the form \cite{Strin1}%

\begin{equation}
\left(  -\frac{1}{2}\overline{\nabla}_{\bot}^{2}+\frac{1}{2}\overline{r}%
_{\bot}^{2}+4\pi an_{1}\left\vert \overline{\varphi}\right\vert ^{2}\right)
\overline{\varphi}=\overline{\mu}_{\bot}(n_{1})\overline{\varphi
}.\label{II-14}%
\end{equation}
This equation, that depends on the sole parameter $an_{1}$, can be
analytically solved in two limiting cases. When $an_{1}\ll1$ the mean-field
interaction energy can be treated as a weak perturbation. In this perturbative
regime the condensate wave function that minimizes the energy functional is
given, to the lowest order, by the Gaussian ground state of the harmonic
oscillator and the condensate is tightly confined in the radial direction.
Under these conditions the radial motion is frozen out, restricted to
zero-point oscillations, and the condensate becomes effectively one
dimensional. Thus the perturbative regime corresponds to quasi-1D mean-field
condensates with a local chemical potential given by%
\begin{equation}
\overline{\mu}_{\bot}(n_{1})=1+2an_{1}.\label{II-15}%
\end{equation}
When $an_{1}\gg1$ the kinetic energy term can be safely neglected in
comparison with the mean-field interaction energy. This is the Thomas-Fermi
regime, in which many modes of the transverse harmonic trap become excited and
the condensate exhibits a parabolic radial profile%
\begin{equation}
\left\vert \overline{\varphi}\right\vert ^{2}=\frac{1}{4\pi an_{1}}\left(
\overline{\mu}_{\bot}-\frac{1}{2}\overline{r}_{\bot}^{2}\right)
,\label{II-16}%
\end{equation}
where the local chemical potential ensuring normalization is now given by%
\begin{equation}
\overline{\mu}_{\bot}(n_{1})=2\sqrt{an_{1}}.\label{II-17}%
\end{equation}
Note that in passing from the Thomas-Fermi regime to the perturbative regime
the system undergoes an effective dimensional crossover from a 3D cigar-shaped
condensate to a quasi-1D mean-field condensate.

In previous works \cite{Previos}, by using a suitable approximation scheme, we
derived general approximate formulas that provide with remarkable accuracy
(typically better than $1\%$) the ground-state properties of any mean-field
scalar Bose-Einstein condensate with short-range repulsive interatomic
interactions, confined in arbitrary cylindrically symmetric harmonic traps,
and even containing a multiply quantized axisymmetric vortex. This
approximation scheme essentially represents an extension of the Thomas-Fermi
approximation that incorporates conveniently the zero-point energy
contribution. In the limiting cases of cigar-shaped and disk-shaped
condensates the ground-state properties follow from explicit analytical
formulas that reduce to the correct analytical expressions in both the TF and
the perturbative regimes, and remain valid and accurate in between these two
limiting cases, thus accounting properly for the corresponding dimensional
crossover. Physical quantities such as the condensate radius, axial length,
chemical potential, mean-field interaction energy, kinetic and potential
energies, density profiles, and local sound velocities can be easily and
accurately obtained in this way in terms of the physically relevant parameters
(number of atoms, trap aspect ratio, and vortex charge). In particular, in
Ref. \cite{Previos}, we found that the transverse local chemical potential of
a condensate with an axisymmetric vortex of charge $q$, as a function of the
condensate density per unit length $n_{1}$, is given by%
\begin{equation}
\overline{\mu}_{\bot}(n_{1})=(|q|+1)+\sqrt{\beta_{q}^{2}+4an_{1}}-\beta
_{q},\label{II-18}%
\end{equation}
with%
\begin{equation}
\beta_{q}=\frac{2^{2|q|}(|q|!)^{2}}{(2|q|)!}.\label{II-19}%
\end{equation}
The parameter $\beta_{q}^{-1}$ accounts for the dilution effect that the
centrifugal force associated with the vortex has on the condensate mean
density \cite{Previos}. In the absence of vortices $q=0\rightarrow\beta_{q}%
=1$, the equation above simplifies to%
\begin{equation}
\overline{\mu}_{\bot}(n_{1})=\sqrt{1+4an_{1}}.\label{II-20}%
\end{equation}
Clearly, in the appropriate limits, this equation reduces to the quasi-1D and
TF expressions (\ref{II-15}) and (\ref{II-17}), respectively. As we shall see,
it also describes correctly the corresponding dimensional crossover.

The above equations were derived in\ Ref. \cite{Previos} by using a suitable
TF-like ansatz for the local density $\left\vert \psi(r_{\bot},z)\right\vert
^{2}$ of a harmonically trapped BEC in its ground state. Such local density,
which is not factorizable in general, is defined in a volume that corresponds
to the usual TF ellipsoidal density cloud conveniently truncated in order to
account for the zero-point energy contribution. Note that even though the
expression (\ref{II-18}) was originally obtained for a condensate in its
ground state (compatible with an axisymmetric vortex of charge $q$) and
axially confined by a harmonic potential, it is, however, of general validity.
This is so because, as already said, the solution of the transverse equation
(\ref{II-14}) does not depend on the particular axial evolution. To convince
the reader that this is the case we have numerically solved Eq. (\ref{II-14})
for a wave function of the form%
\begin{equation}
\overline{\varphi}(\overline{\mathbf{r}}_{\bot})=\exp(iq\theta)\overline
{\varphi}_{q}(\overline{r}_{\bot}),\label{II-21}%
\end{equation}
with $q=0,1,2,$ and $4$. The results of the numerical calculation are shown in
Fig. \ref{Fig1} (open circles) along with the theoretical prediction obtained
from Eq. (\ref{II-18}) (solid lines). As is apparent, Eq. (\ref{II-18})
accurately accounts for the dimensional crossover, the agreement with the
numerical results being excellent for any value of the dimensionless
interaction parameter $an_{1}$. The maximum error is smaller than $1\%$ for
$q=0$ and $2$, and smaller than $1.2\%$ for $q=1$. Even though one expects
\cite{Previos} this maximum error to increase with $q$, it is still smaller
than $2.5\%$ for $q=4$. This demonstrates that the local chemical potential as
given by Eq. (\ref{II-18}) is an accurate solution of the transverse equation
(\ref{II-8}). This result is of interest in its own right. For instance, in
Ref. \cite{Kramer1}, it has been shown that the velocity of sound in an
axially uniform Bose-Einstein condensate immersed in a 1D optical lattice and
radially confined by a harmonic trap can be obtained, in a wide range of
optical lattice depths, from the knowledge of the local chemical potential
$\overline{\mu}_{\bot}$ as a function of the linear average density
$n_{1}=N/d$, with $d$ being the lattice period. In such an approach the effect
of the lattice is incorporated into a suitable renormalization of the mass and
of the coupling constant. In the present work, however, the main interest of
Eq. (\ref{II-18}) comes from the fact that when substituted in Eq.
(\ref{II-6}) it leads to the following effective 1D mean-field equation:%
\begin{figure}
[ptb]
\begin{center}
\includegraphics[
height=6.7121cm,
width=7.7939cm
]%
{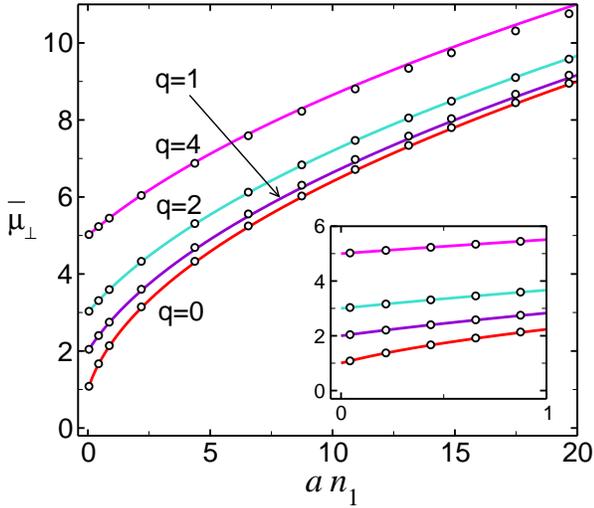}%
\caption{(Color online) Theoretical prediction for the transverse local
chemical potential $\overline{\mu}_{\bot}$ as a function of $an_{1}$ for
different vortex charges $q$ (solid lines). The open circles are exact
numerical results obtained by solving the stationary Gross-Pitaevskii equation
(\ref{II-14}).}%
\label{Fig1}%
\end{center}
\end{figure}
%

\begin{equation}
i\hbar\frac{\partial\phi}{\partial t}=-\frac{\hbar^{2}}{2m}\frac{\partial
^{2}\phi}{\partial z^{2}}+V_{z}(z)\phi+\hbar\omega_{\bot}\sqrt{\beta_{q}%
^{2}+4aN\left\vert \phi\right\vert ^{2}}\phi,\label{II-22}%
\end{equation}
where $\beta_{q}$ is given by Eq. (\ref{II-19}) and we have absorbed the
constant term proportional to $(|q|+1)-\beta_{q}$ into the definition of the
generic axial potential $V_{z}(z)$. For $q=0,1,2,...$ the parameter $\beta
_{q}$ takes the values $1,2,8/3,...$, respectively.\ 

Equation (\ref{II-22}) is the equation we were looking for. It governs the
axial dynamics of arbitrary mean-field cigar-shaped condensates with repulsive
interatomic interactions even in the presence of an axisymmetric vortex of
charge $q$, accounting for the effects from the transverse degrees of freedom
through the term proportional to $\hbar\omega_{\bot}$. Since vortices with
$q\geq2$ are dynamically unstable and decay into an array of singly quantized
vortices \cite{PRL06,Mott}, in such cases the applicability of Eq.
(\ref{II-22}) is restricted to times shorter than the corresponding decay
time. As is apparent, the equation above is much simpler than previous
proposals. It is also more accurate, as we shall see. Since we have found an
expression for the transverse local chemical potential that is very accurate,
the validity of Eq. (\ref{II-22}) relies almost exclusively on the validity of
the well established adiabatic approximation.

When $4an_{1}\ll\beta_{q}^{2}$ one enters the quasi-1D mean-field regime. In
this case, Eq. (\ref{II-22}) reduces to%
\begin{equation}
i\hbar\frac{\partial\phi}{\partial t}=-\frac{\hbar^{2}}{2m}\frac{\partial
^{2}\phi}{\partial z^{2}}+V_{z}(z)\phi+g_{\mathrm{1D}}N\left\vert
\phi\right\vert ^{2}\phi,\label{II-23}%
\end{equation}
with%
\begin{equation}
g_{\mathrm{1D}}=\beta_{q}^{-1}\frac{g}{2\pi a_{\bot}^{2}}=\beta_{q}%
^{-1}2a\hbar\omega_{\bot}.\label{II-24}%
\end{equation}
This equation generalizes the well-known 1D GPE to the case of condensates
containing an axisymmetric vortex. In the absence of vortices, $\beta_{q}=1$
and one recovers the usual equation. When the number of particles is high
enough that $4an_{1}\gg\beta_{q}^{2}$, the condensate enters the TF regime. In
this case, Eq. (\ref{II-22}) reduces to%
\begin{equation}
i\hbar\frac{\partial\phi}{\partial t}=-\frac{\hbar^{2}}{2m}\frac{\partial
^{2}\phi}{\partial z^{2}}+V_{z}(z)\phi+2\hbar\omega_{\bot}\sqrt{aN}\left\vert
\phi\right\vert \phi,\label{II-25}%
\end{equation}
which again is the correct result as follows from the direct substitution of
Eq. (\ref{II-17}) into Eq. (\ref{II-6}).

The effective 1D equation (\ref{II-22}) also permits deriving an accurate
analytical expression for the stationary axial density profile $n_{1}%
(z)=N|\phi_{0}(z)|^{2}$. To see this we shall consider a harmonic axial
confinement $V_{z}(z)=\frac{1}{2}m\omega_{z}^{2}z^{2}$ with corresponding
oscillator length $a_{z}=\sqrt{\hbar/m\omega_{z}}$. For cigar-shaped
condensates the trap aspect ratio $\lambda=\omega_{z}/\omega_{\bot}$ must
satisfy the inequality $\lambda\ll1$. Under these circumstances the condition
(\ref{II-11}) can be easily fulfilled. It would be sufficient (though not
necessary) that $\Delta_{z}\gg a_{\bot}/\sqrt{4an_{1}}$. Taking into account
that, in the stationary state, $\Delta_{z}$ is of the order of the condensate
axial length it is clear that, except for extremely small values of $an_{1}$,
in cigar-shaped condensates this condition always holds. As a consequence, the
stationary equation (\ref{II-10}) reduces to the local density approximation
(\ref{II-12}). Substitution of Eq. (\ref{II-18}) into Eq. (\ref{II-12}) yields%
\begin{equation}
4an_{1}(\overline{z})=\left(  \frac{\mu}{\hbar\omega_{\bot}}-(|q|+1)+\beta
_{q}-\frac{1}{2}(\sqrt{\lambda}\overline{z})^{2}\right)  ^{2}-\beta_{q}%
^{2},\label{II-26}%
\end{equation}
where $\overline{z}=z/a_{z}$. The dimensionless axial half-length
$\overline{Z}=Z/a_{z}$ follows from the condition $n_{1}(\overline{Z})=0$%
\begin{equation}
\frac{\mu}{\hbar\omega_{\bot}}=(|q|+1)+\frac{1}{2}(\sqrt{\lambda}%
\,\overline{Z})^{2}.\label{II-27}%
\end{equation}
Substituting this expression in Eq. (\ref{II-26}) we obtain%
\begin{equation}
n_{1}(z)=\beta_{q}\frac{(\sqrt{\lambda}\;\overline{Z})^{2}}{4a}\left(
1-\frac{z^{2}}{Z^{2}}\right)  +\frac{(\sqrt{\lambda}\;\overline{Z})^{4}}%
{16a}\left(  1-\frac{z^{2}}{Z^{2}}\right)  ^{2}\label{II-28}%
\end{equation}
with $n_{1}(z)=0$ for $|z|>Z$. In order for this equation to be useful one
also needs an analytic expression for $\overline{Z}$ or, equivalently, for
$\mu$. From the normalization condition%
\begin{equation}
N=\int_{-Z}^{+Z}dz\,n_{1}(z),\label{II-29}%
\end{equation}
one finds that the axial half-length satisfies the quintic polynomial equation%
\begin{equation}
\frac{1}{15}(\sqrt{\lambda}\,\overline{Z})^{5}+\frac{1}{3}\beta_{q}%
(\sqrt{\lambda}\,\overline{Z})^{3}=\chi_{1},\label{II-30}%
\end{equation}
where $\chi_{1}\equiv\lambda Na/a_{\bot}$ is (apart from the vortex charge)
the only relevant parameter. An approximate solution of the above equation is
given by%
\begin{equation}
\sqrt{\lambda}\,\overline{Z}=\left[  \frac{1}{\left(  15\chi_{1}\right)
^{\frac{4}{5}}+\frac{1}{3}}+\frac{1}{57\chi_{1}+345}+\frac{1}{(3\chi_{1}%
/\beta_{q})^{\frac{4}{3}}}\right]  ^{-\frac{1}{4}}\label{II-31}%
\end{equation}
This expression satisfies Eq. (\ref{II-30}) for any $\chi_{1}\in
\lbrack0,\infty)$, with a residual error \cite{Previos} that is smaller than
$0.75\%$ for $q=0$\ and smaller than $3.2\%$ for $1\leq|q|\leq10$. The (local)
axial (first) sound velocity $c_{\mathrm{1D}}$ of a cigar-shaped condensate,
is defined by%
\begin{equation}
c_{\mathrm{1D}}^{2}=\frac{n_{1}}{m}\frac{\partial\mu_{\bot}}{\partial n_{1}%
}.\label{II-32}%
\end{equation}
Substituting Eq. (\ref{II-18}) in Eq. (\ref{II-32}) one obtains%
\begin{equation}
\frac{mc_{\mathrm{1D}}^{2}}{\hbar\omega_{\bot}}=\sqrt{\frac{4a^{2}n_{1}%
^{2}(z)}{\beta_{q}^{2}+4an_{1}(z)}}.\label{II-33}%
\end{equation}
Equations (\ref{II-27})--(\ref{II-33}) coincide with the results we obtained
in previous works \cite{Previos}. Note, however, that the derivation is quite
different. The formulation of Ref. \cite{Previos} is not restricted to
condensates of specific geometry. The formulas obtained there are valid for
condensates confined in arbitrary cylindrically symmetric harmonic traps and
reduce to those derived in the present work\ in the limiting case of highly
elongated condensates. In particular, Eq. (\ref{II-27}) above appears in Ref.
\cite{Previos}\ as part of the initial ansatz for the local density
$\left\vert \psi(r_{\bot},z)\right\vert ^{2}$. On the other hand, Eq.
(\ref{II-28}) follows in Ref. \cite{Previos} from a direct integration of this
ansatz over the radial coordinate, while Eq. (\ref{II-30}) is obtained after
integrating $\left\vert \psi(r_{\bot},z)\right\vert ^{2}$ over both the radial
and the axial coordinates.

As shown in Ref. \cite{Previos}, Eqs (\ref{II-27})--(\ref{II-33}) predict very
accurately the condensate ground-state properties. They also reduce to the
correct expressions in the two analytically solvable regimes. In particular,
in the quasi-1D mean-field regime (corresponding to $\chi_{1}\ll1\rightarrow
an_{1}\ll1$) the axial density profile (\ref{II-28}) is well approximated by
the first term on the right-hand side while, in the TF regime (corresponding
to $\chi_{1}\gg1\rightarrow an_{1}\gg1$), the last term on the right-hand side
is the only one that contributes significantly, in good agreement with
previous results obtained in these two particular limits \cite{Strin1}.
Equation (\ref{II-28}) also reproduces accurately the axial density profile in
between these two limiting cases. This has been demonstrated in Ref.
\cite{Previos}, where we compare the theoretical prediction obtained from Eqs.
(\ref{II-28}) and (\ref{II-31}) with exact numerical results obtained from the
full 3D GPE. The agreement is always very good except at the condensate edges,
as expected for a TF-like expression obtained by neglecting the derivative
kinetic term. In this regard, we note again that the derivation of the above
equations followed in the present work is based on the local density
approximation which in turn requires the condition (\ref{II-11}) to be true.
As already said, however, for cigar-shaped condensates with $\lambda\ll1$ this
requirement is not very demanding and can be satisfied even in the
perturbative regime (see also Ref. \cite{Strin1}).

Our effective 1D equation (\ref{II-22}) is a Schr\"{o}dinger equation with an
effective mean-field potential%
\begin{equation}
V_{\mathrm{eff}}(\phi)=\hbar\omega_{\bot}\sqrt{\beta_{q}^{2}+4aN\left\vert
\phi\right\vert ^{2}}.\label{II-34}%
\end{equation}
As mentioned in the Introduction, by using a variational approach
\cite{Victor1}, Salasnich \emph{et al.} obtained in Ref. \cite{Reatto1}
effective 1D and 2D nonpolynomial nonlinear Schr\"{o}dinger equations (NPSEs)
for the axial and radial dynamics of cigar-shaped and disk-shaped condensates,
respectively.\ Their 1D equation coincides with our equation above but with a
different effective mean-field potential%
\begin{align}
V_{\mathrm{eff}}(\phi) &  =\frac{gN}{2\pi a_{\bot}^{2}}\frac{\left\vert
\phi\right\vert ^{2}}{\sqrt{1+2aN\left\vert \phi\right\vert ^{2}}}\nonumber\\
&  +\frac{\hbar\omega_{\bot}}{2}\left(  \frac{1}{\sqrt{1+2aN\left\vert
\phi\right\vert ^{2}}}+\sqrt{1+2aN\left\vert \phi\right\vert ^{2}}\right)
.\label{II-35}%
\end{align}
Next we will compare our effective 1D equation (\ref{II-22}) with that by
Salasnich \emph{et al.} These authors demonstrated that their equations
provide much more accurate results than any other proposed effective equation.
Actually, their results are always very close to the exact results obtained
from the 3D GPE. It is thus sufficient to compare our equations with those
proposed in Ref. \cite{Reatto1}. As we shall see, our effective equation is
more accurate. Moreover, it allows treating condensates containing an
axisymmetric vortex, with no additional cost. For instance, to account for a
$q=1$ vortex it suffices to make the change $\beta_{q=0}=1\rightarrow
\beta_{q=1}=2$ in Eq. (\ref{II-34}). Our effective equation also permits one
to derive useful and accurate analytical expressions for ground-state
properties such as the chemical potential, axial length, axial density
profile, and local sound velocity. In this regard, it should be noticed that
although Salasnich \emph{et al.} also obtained in Ref. \cite{Reatto2}
analytical expressions for the axial density profile and local sound velocity,
such expressions depend on the condensate chemical potential which because of
the complexity of Eq. (\ref{II-35}) in general cannot be determined
analytically except in the two analytically solvable regimes or in the simple
case of a homogeneous condensate with no axial confinement. Finally, note that
the approach of Ref. \cite{Reatto1} is also valid for condensates with
attractive interatomic interactions.%
\begin{figure}
[ptb]
\begin{center}
\includegraphics[
height=9.2686cm,
width=8.4102cm
]%
{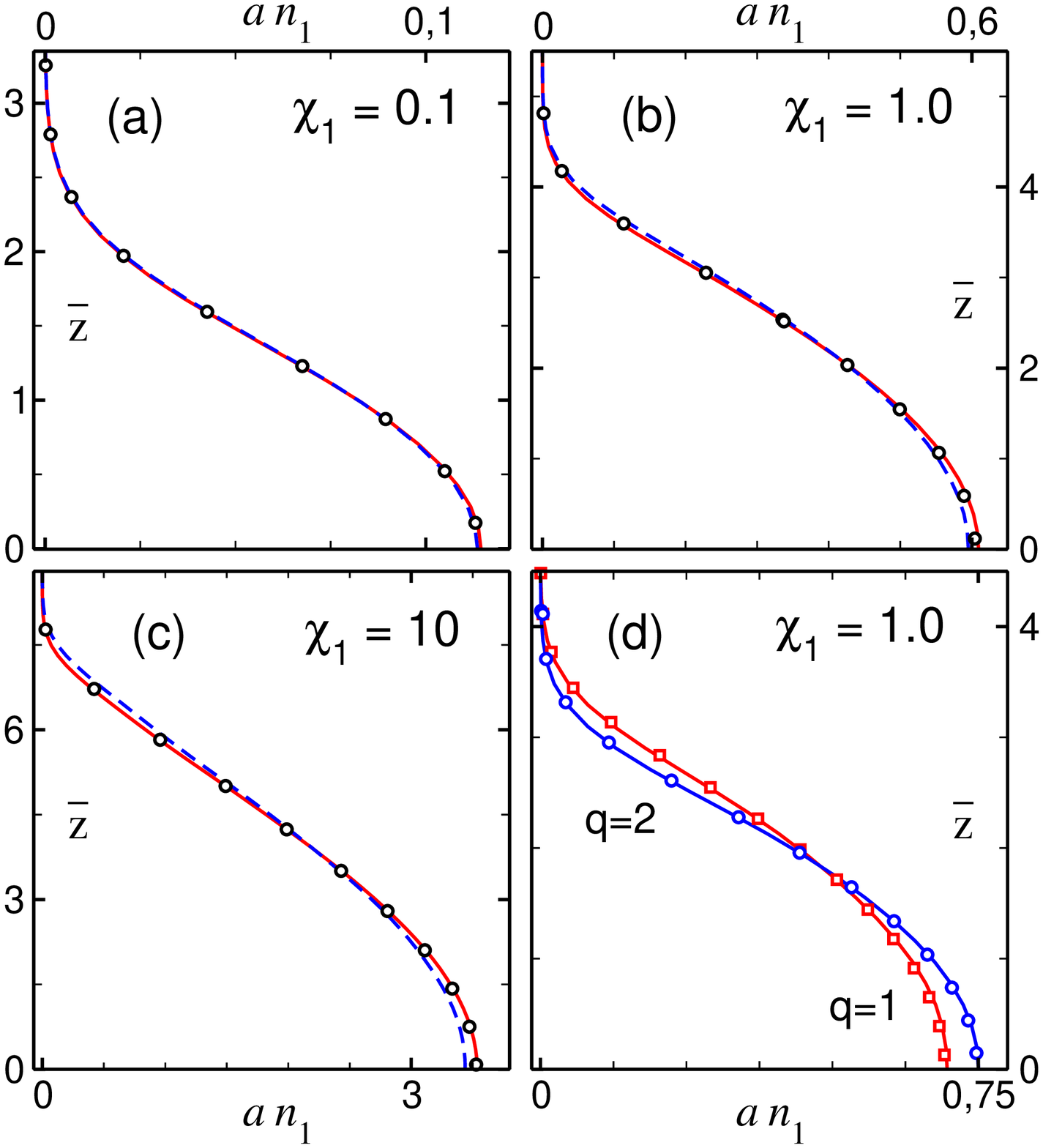}%
\caption{(Color online) Axial density profile $n_{1}(\overline{z})$ of
cigar-shaped condensates in a harmonic trap with aspect ratio $\lambda
=\omega_{z}/\omega_{\bot}=0.1$ and $\chi_{1}=0.1,1$ and $10$. (a)-(c):
Ground-state ($q=0$) equilibrium configuration. (d): Equilibrium configuration
compatible with a vortex of charge $q=1$ and $2$. Solid lines are numerical
results obtained from our effective 1D equation (\ref{II-22}). Dashed lines
are numerical results obtained from the 1D NPSE by Salasnich \emph{et al.}
Open symbols are exact numerical results obtained from the full 3D GPE.}%
\label{Fig3}%
\end{center}
\end{figure}

Figure \ref{Fig3} shows the axial density profile $n_{1}(\overline{z})$ of
cigar-shaped condensates in a harmonic trap with aspect ratio $\lambda
=\omega_{z}/\omega_{\bot}=0.1$ in the different relevant regimes, which, for a
given vorticity $q$, are completely characterized by the value of the sole
parameter $\chi_{1}$. Figures \ref{Fig3}(a)--\ref{Fig3}(c) correspond to the
ground-state ($q=0$) equilibrium configuration, while Fig. \ref{Fig3}(d) shows
the equilibrium configuration compatible with a vortex of charge $q=1$ and
$2$. Solid lines are numerical results obtained from our effective 1D equation
(\ref{II-22}). Dashed lines are numerical results obtained from the 1D NPSE by
Salasnich \emph{et al.}\ and open symbols are exact numerical results obtained
from the full 3D GPE. As is apparent from Fig. \ref{Fig3}(a), in the
perturbative regime ($\chi_{1}=0.1$), both approaches are practically
indistinguishable, a consequence of the fact that both have the correct
perturbative limit. However, as $\chi_{1}$ increases our effective 1D equation
is always more accurate [Figs. \ref{Fig3}(b) and \ref{Fig3}(c)], which is a
consequence of the fact that our approach, unlike that by Salasnich \emph{et
al.}, also reproduces correctly the TF limit. The difference, though small,
can be clearly appreciated even for $\chi_{1}=1$. Moreover, our approach also
allows treating the case of condensates containing an axisymmetric vortex,
with no additional cost. Figure \ref{Fig3}(d) shows the equilibrium
configuration of condensates with a vortex of charge $q=1$ and $2$, obtained
from the same effective 1D equation (\ref{II-22}) by simply taking $\beta
_{q}=2$ and $8/3$, respectively.\ As is apparent, the agreement with the exact
numerical results (open symbols) is again very good.%
\begin{figure}
[ptb]
\begin{center}
\includegraphics[
height=9.3316cm,
width=8.4102cm
]%
{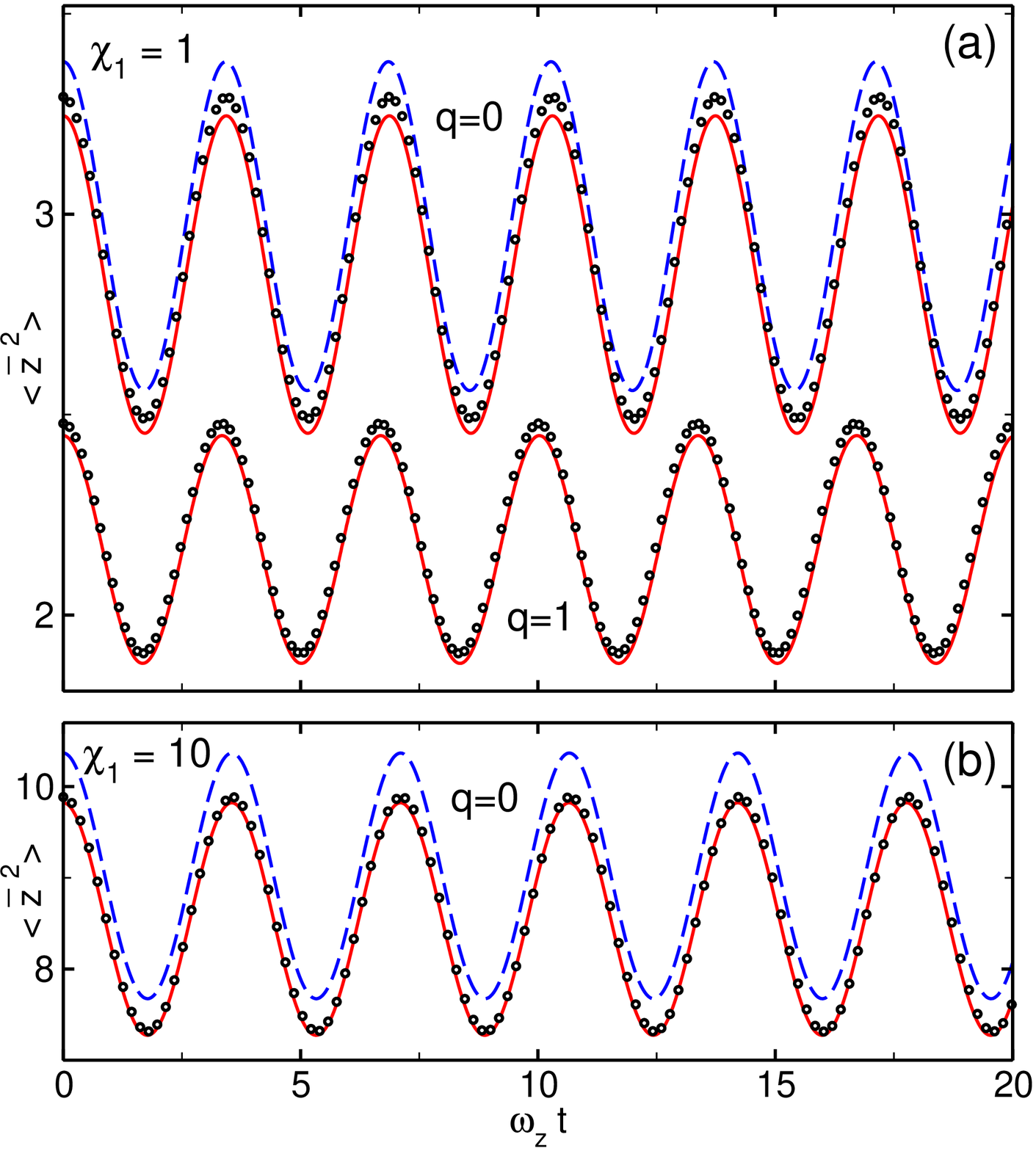}%
\caption{(Color online) Time evolution of the mean squared amplitude
$\langle\overline{z}^{2}\rangle$ of cigar-shaped condensates with $\chi_{1}=1$
and $10$ and vortex charges $q=0$ and $1$ after a perturbation $\omega
_{z}\rightarrow1.1\omega_{z}$. The aspect ratio of the harmonic trap before
perturbation is $\lambda=\omega_{z}/\omega_{\bot}=0.1$. Solid lines are
numerical results obtained from our effective 1D equation (\ref{II-22}).
Dashed lines are numerical results obtained from the 1D NPSE by Salasnich
\emph{et al.} Open symbols are exact numerical results obtained from the full
3D GPE.}%
\label{Fig4}%
\end{center}
\end{figure}

Our approach also reproduces very accurately the time evolution of
cigar-shaped condensates. To see this we start from the equilibrium
configuration of condensates confined in a harmonic trap with aspect ratio
$\lambda=\omega_{z}/\omega_{\bot}=0.1$. Then we introduce a sudden
perturbation by increasing the axial confinement frequency as $\omega
_{z}\rightarrow1.1\omega_{z}$ and follow the subsequent evolution in time of
the mean squared axial amplitude $\langle\overline{z}^{2}\rangle=\int
d\overline{z}\,\overline{z}^{2}n_{1}(\overline{z},t)$. Figure \ref{Fig4} shows
the corresponding results obtained for condensates with $\chi_{1}=1$ and $10$
and vortex charges $q=0$ and $1$. As before solid lines are numerical results
obtained from our effective 1D equation (\ref{II-22}), dashed lines are
numerical results obtained from the 1D NPSE by Salasnich \emph{et al.}, and
open symbols are exact numerical results obtained from the full 3D GPE. For
$\chi_{1}=0.1$ (not shown in the figure) both approaches produce results that
are indistinguishable on the scale of the figure from the exact results. It is
apparent that our approach is more accurate than that by Salasnich \emph{et
al.} Moreover, as Fig. \ref{Fig4}(a) shows, it also produces very accurate
results for condensates containing an axisymmetric vortex. Note also that for
$\chi_{1}=1$ and $q=0$, Fig. \ref{Fig4}(a) shows that the results from the 1D
NPSE begin to dephase at large times. To see this behavior more clearly, it is
convenient to consider a more complex situation. To this end we consider a BEC
with $\chi_{1}=2$ in a harmonic trap with $\lambda=0.1$, subject to a
blue-detuned laser beam modelled by the Gaussian potential%
\begin{equation}
V(z)=V_{0}\exp(-z^{2}/2z_{0}^{2}),\label{II-36}%
\end{equation}%
\begin{figure}
[ptb]
\begin{center}
\includegraphics[
height=9.2279cm,
width=8.4097cm
]%
{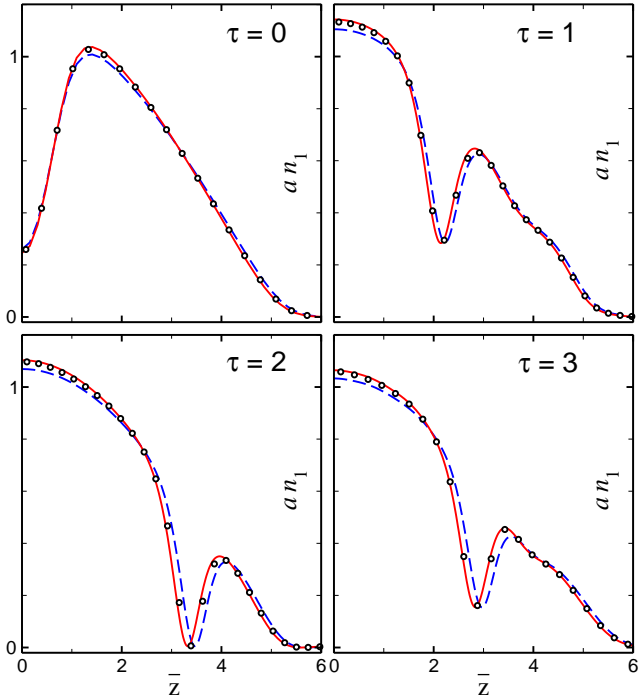}%
\caption{(Color online) Evolution of\ the axial density profile $n_{1}%
(\overline{z})$ of a cigar-shaped condensate in a harmonic trap with aspect
ratio $\lambda=\omega_{z}/\omega_{\bot}=0.1$ and $\chi_{1}=2$ for different
times $\tau=\omega_{z}t$. Solid lines are numerical results obtained from our
effective 1D equation (\ref{II-22}). Dashed lines are numerical results
obtained from the 1D NPSE by Salasnich \emph{et al.} Open symbols are exact
numerical results obtained from the full 3D GPE.}%
\label{Fig5}%
\end{center}
\end{figure}
with $V_{0}=12\hbar\omega_{z}$ and $z_{0}=\sqrt{2\lambda}a_{z}$. We let the
condensate reach its equilibrium configuration and then, at $\tau=\omega
_{z}t=0$, we suddenly switch off the laser beam and let the system evolve in
the harmonic trap. Figure \ref{Fig5} shows the time evolution of the axial
density profile $n_{1}(\overline{z})$. Solid lines are numerical results
obtained from our effective 1D equation (\ref{II-22}), dashed lines are
numerical results obtained from the 1D NPSE by Salasnich \emph{et al.}, and
open symbols are exact numerical results obtained from the full 3D GPE. At
$\tau=1$ one recognizes in each half-axis a dark solitonlike structure that
develops from the initial configuration and propagates toward the
corresponding condensate edge (only the positive half-axis is represented in
the figure). At $\tau=2$ the soliton becomes black and it comes back toward
the condensate center for $\tau>2$, as can be seen in the snapshot at $\tau
=3$. This figure demonstrates again that our approach is more accurate than
that by Salasnich \emph{et al.} In particular, it is apparent that the results
obtained from the 1D NPSE become somewhat dephased with respect to the exact results.

\section{III. DISK-SHAPED CONDENSATES}

Consider now a condensate confined in an anisotropic trap that is much
stronger in the axial than in the radial direction. In this case, using
similar arguments as before, one can resort to the adiabatic approximation and
assume that at every instant of time the (fast) axial degrees of freedom
adjust instantaneously to the equilibrium configuration compatible with the
radial configuration occurring at that time. Under these circumstances the
condensate wave function can be factorized as%
\begin{equation}
\psi(\mathbf{r},t)=\varphi(\mathbf{r}_{\bot},t)\phi(z;n_{2}(\mathbf{r}_{\bot
},t)),\label{III-1}%
\end{equation}
where $n_{2}(\mathbf{r}_{\bot},t)$ is the local condensate density per unit
area characterizing the radial configuration%
\begin{equation}
n_{2}(\mathbf{r}_{\bot},t)\equiv N\int dz|\psi(\mathbf{r}_{\bot}%
,z,t)|^{2}.\label{III-2}%
\end{equation}
The normalization condition%
\begin{equation}
\int dz|\phi(z;n_{2})|^{2}=1\label{III-3}%
\end{equation}
leads to%
\begin{equation}
n_{2}(\mathbf{r}_{\bot},t)=N|\varphi(\mathbf{r}_{\bot},t)|^{2}.\label{III-4}%
\end{equation}
After substituting Eq. (\ref{III-1}) into Eq. (\ref{I-1}), one arrives at%
\begin{multline}
\left(  i\hbar\frac{\partial\varphi}{\partial t}+\frac{\hbar^{2}}{2m}%
\nabla_{\bot}^{2}\varphi-V_{\bot}(\mathbf{r}_{\bot})\varphi\right)
\phi(z;n_{2})=\\
\left(  -\frac{\hbar^{2}}{2m}\frac{\partial^{2}\phi}{\partial z^{2}}%
+V_{z}(z)\phi+gn_{2}(\mathbf{r}_{\bot},t)\left\vert \phi\right\vert ^{2}%
\phi\right)  \varphi(\mathbf{r}_{\bot},t),\label{III-5}%
\end{multline}
where as before we have assumed the confining potential to be separable.
Integrating out the fast degrees of freedom one obtains%
\begin{equation}
i\hbar\frac{\partial\varphi}{\partial t}=-\frac{\hbar^{2}}{2m}\nabla_{\bot
}^{2}\varphi+V_{\bot}(\mathbf{r}_{\bot})\varphi+\mu_{z}(n_{2})\varphi
,\label{III-6}%
\end{equation}
with%
\begin{equation}
\mu_{z}(n_{2})\equiv\int dz\,\phi^{\ast}\left(  -\frac{\hbar^{2}}{2m}%
\frac{\partial^{2}}{\partial z^{2}}+V_{z}(z)+gn_{2}\left\vert \phi\right\vert
^{2}\right)  \phi.\label{III-7}%
\end{equation}
Substituting Eq. (\ref{III-6}) in Eq. (\ref{III-5}) one finds%
\begin{equation}
\left(  -\frac{\hbar^{2}}{2m}\frac{\partial^{2}}{\partial z^{2}}%
+V_{z}(z)+gn_{2}\left\vert \phi\right\vert ^{2}\right)  \phi=\mu_{z}%
(n_{2})\phi.\label{III-8}%
\end{equation}
The effective 2D mean-field equation (\ref{III-6}) governs the radial dynamics
of disk-shaped BECs, incorporating the contribution from the axial degrees of
freedom through the (axial) local chemical potential $\mu_{z}(n_{2})$. This
quantity, in turn, follows from the \emph{stationary} Gross-Pitaevskii
equation (\ref{III-8}) determining the axial configuration $\phi(z;n_{2})$.

The condensate stationary states%
\begin{equation}
\varphi(\mathbf{r}_{\bot},t)=\varphi_{0}(\mathbf{r}_{\bot})e^{-i\mu t/\hbar
}\label{III-9}%
\end{equation}
satisfy%
\begin{equation}
-\frac{\hbar^{2}}{2m}\nabla_{\bot}^{2}\varphi_{0}+V_{\bot}(\mathbf{r}_{\bot
})\varphi_{0}+\mu_{z}(n_{2})\varphi_{0}=\mu\varphi_{0},\label{III-10}%
\end{equation}
where the chemical potential$\ \mu$ guarantees the condition $\int
d^{2}\mathbf{r}_{\bot}|\varphi(\mathbf{r}_{\bot},t)|^{2}=1$. When the typical
length scale $\Delta_{\bot}$ of the spatial variations of $\varphi_{0}$ is
much greater than the corresponding \emph{radial healing length}, i.e.,%
\begin{equation}
\Delta_{\bot}\gg\frac{\hbar}{\sqrt{2m(\mu_{z}(n_{2})-\frac{1}{2}\hbar
\omega_{z})}},\label{III-11}%
\end{equation}
then Eq. (\ref{III-10}) reduces to the local density approximation%
\begin{equation}
\mu=\mu_{z}(n_{2})+V_{\bot}(\mathbf{r}_{\bot}).\label{III-12}%
\end{equation}
We shall consider in what follows disk-shaped condensates confined in the
axial direction by a harmonic potential characterized by an oscillator length
$a_{z}=\sqrt{\hbar/m\omega_{z}}$,%
\begin{equation}
V(\mathbf{r})=\frac{1}{2}m\omega_{z}^{2}z^{2}+V_{\bot}(\mathbf{r}_{\bot
}).\label{III-13}%
\end{equation}
In terms of dimensionless variables $\overline{z}=z/a_{z}$, $\overline{\phi
}=\sqrt{a_{z}}\phi$, and $\overline{\mu}_{z}=\mu_{z}/\hbar\omega_{z}$, the
axial equation (\ref{III-8}) reads%
\begin{equation}
\left(  -\frac{1}{2}\frac{\partial^{2}}{\partial\overline{z}^{2}}+\frac{1}%
{2}\overline{z}^{2}+4\pi aa_{z}n_{2}\left\vert \overline{\phi}\right\vert
^{2}\right)  \overline{\phi}=\overline{\mu}_{z}(n_{2})\overline{\phi
}.\label{III-14}%
\end{equation}
In Ref. \cite{Previos}, we found the following expression for the axial local
chemical potential $\overline{\mu}_{z}$ as a function of the condensate
density per unit area $n_{2}$:%
\begin{equation}
\overline{\mu}_{z}(n_{2})\equiv{}\!\frac{1}{8}\!\left[  \!\left(
\!\eta+\!\sqrt{\eta^{2}\!-\!\xi_{1}^{6}}\right)  ^{\frac{1}{3}}\!+\!\left(
\!\eta\!-\!\sqrt{\eta^{2}\!-\!\xi_{1}^{6}}\right)  ^{\frac{1}{3}}%
\!\!-\!\xi_{1}\right]  ^{2},\label{III-15}%
\end{equation}
where $\eta=4+6\xi_{1}-\xi_{1}^{3}+24\pi aa_{z}n_{2}$ and $\xi_{1}%
\equiv(\kappa_{2}-1)$ with%
\begin{align}
\kappa_{2}^{-1}(\overline{n}_{2}) &  \equiv\sqrt{2/\pi}+\Theta(\overline
{n}_{2}-0.1)\nonumber\\
\times &  \left(  1-\sqrt{2/\pi}\right)  \left(  1-(10\overline{n}_{2}%
)^{-1/5}\right)  .\label{III-16}%
\end{align}
In the above equation, $\Theta(x)$ is the step function and $\overline{n}%
_{2}\equiv aa_{z}n_{2}$ is the only relevant parameter. Note that in Ref.
\cite{Previos} we used the same expression\ (\ref{III-16})\ but in terms of
the dimensionless parameter $\chi_{2}\equiv Na/\lambda^{2}a_{z}$ instead of
$\overline{n}_{2}$. It was shown there that $\chi_{2}$ is the only relevant
parameter for the description of harmonically trapped disk-shaped condensates,
so that $\overline{n}_{2}$ and $\chi_{2}$ are directly related to each other.
In particular, the quasi-2D perturbative regime corresponds to $\chi_{2}%
\ll1\leftrightarrow\overline{n}_{2}\ll1$, while in the TF regime $\chi_{2}%
\gg1\leftrightarrow\overline{n}_{2}\gg1$. On the other hand, as explained in
Ref. \cite{Previos}, the (slowly varying) last term in Eq. (\ref{III-16}) was
conveniently introduced to ensure that in the TF limit $\kappa_{2}%
^{-1}\rightarrow1$, its specific functional form being not very relevant. This
way one guarantees the correct expression for $\overline{\mu}_{z}(n_{2})$ in
both the TF and the perturbative regimes. On these grounds and taking into
account that $\chi_{2}\sim$ $\overline{n}_{2}$, we have simply made the
substitution $\chi_{2}\rightarrow$ $\overline{n}_{2}$ in Eq. (\ref{III-16})
which now becomes a very slowly varying function of $\overline{n}_{2}$.%
\begin{figure}
[ptb]
\begin{center}
\includegraphics[
height=6.7475cm,
width=7.7929cm
]%
{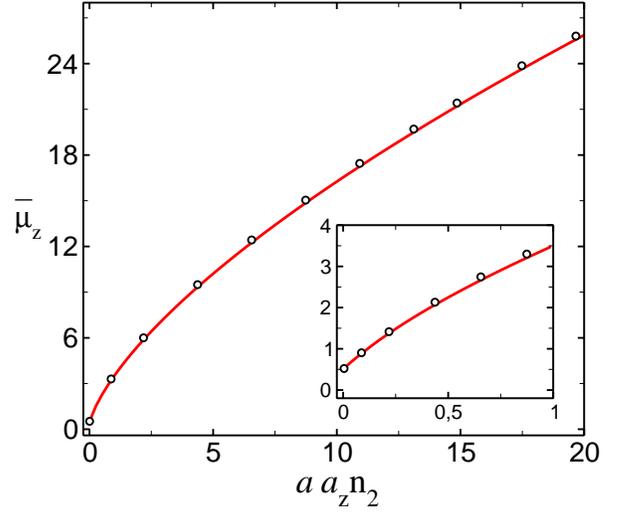}%
\caption{(Color online) Theoretical prediction for the axial local chemical
potential $\overline{\mu}_{z}$ as a function of $aa_{z}n_{2}$ (solid lines).
The open circles are exact numerical results obtained by solving the
stationary Gross-Pitaevskii equation (\ref{III-14}).}%
\label{Fig6}%
\end{center}
\end{figure}

Figure \ref{Fig6} shows the theoretical prediction for $\overline{\mu}_{z}$
obtained from Eq. (\ref{III-15}) (solid lines) along with the exact results
obtained from the numerical solution of the axial equation (\ref{III-14})
(open circles). As is apparent, not only does Eq. (\ref{III-15}) have the
correct limit in the two relevant extreme regimes, but it also accounts
accurately for the corresponding dimensional crossover.

Substituting Eqs. (\ref{III-15}) and (\ref{III-16}) with $n_{2}=N|\varphi
(\mathbf{r}_{\bot},t)|^{2}$ into Eq. (\ref{III-6}), we finally obtain the
desired effective 2D equation. This equation governs the transverse dynamics
of arbitrary mean-field disk-shaped condensates with repulsive interatomic
interactions, accounting properly for the effects from the axial degrees of
freedom. In the quasi-2D mean-field regime ($aa_{z}n_{2}\ll1$), it reduces to%
\begin{equation}
i\hbar\frac{\partial\varphi}{\partial t}=-\frac{\hbar^{2}}{2m}\nabla_{\bot
}^{2}\varphi+V_{\bot}(\mathbf{r}_{\bot})\varphi+g_{\mathrm{2D}}N|\varphi
|^{2}\varphi,\label{III-17}%
\end{equation}
where $g_{\mathrm{2D}}=g/\sqrt{2\pi}\,a_{z}=2\sqrt{2\pi}aa_{z}\hbar\omega_{z}$
and we have absorbed a constant $\hbar\omega_{z}/2$ into the definition of the
generic transverse potential $V_{\bot}(\mathbf{r}_{\bot})$. Similarly, in the
TF regime ($aa_{z}n_{2}\gg1$), the effective 2D equation reduces to
\begin{align}
i\hbar\frac{\partial\varphi}{\partial t} &  =-\frac{\hbar^{2}}{2m}\nabla
_{\bot}^{2}\varphi+V_{\bot}(\mathbf{r}_{\bot})\varphi\nonumber\\
&  +\hbar\omega_{z}\left[  (3\pi/\sqrt{2})aa_{z}N|\varphi|^{2}\right]
^{2/3}\varphi.\label{III-18}%
\end{align}%
\begin{figure}
[ptb]
\begin{center}
\includegraphics[
height=9.0887cm,
width=8.4097cm
]%
{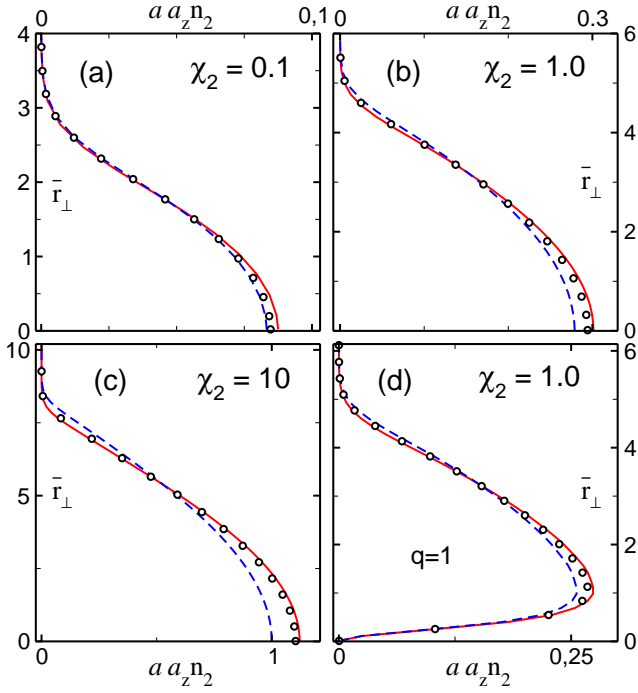}%
\caption{(Color online) Radial density profile $n_{2}(r_{\bot})$ of
disk-shaped condensates in a harmonic trap with aspect ratio $\lambda
=\omega_{z}/\omega_{\bot}=10$ and $\chi_{2}=0.1,1,$ and $10$. (a)-(c):
Ground-state ($q=0$) equilibrium configuration. (d): Equilibrium configuration
compatible with a vortex of charge $q=1$. Solid lines are numerical results
obtained from our effective 2D equation (\ref{III-6}). Dashed lines are
numerical results obtained from the 2D NPSE by Salasnich \emph{et al.} Open
symbols are exact numerical results obtained from the full 3D GPE.}%
\label{Fig7}%
\end{center}
\end{figure}
It can be easily verified that both Eq. (\ref{III-17}) and Eq. (\ref{III-18})
are the correct limits of the underlying 3D GPE. As in the 1D case, our
effective 2D equation permits one to derive analytical expressions for
ground-state properties of disk-shaped condensates such as the chemical
potential, condensate radius, radial density profile, and local sound
velocity. These analytical expressions reduce to the correct formulas in both
the TF and the perturbative regimes, and remain valid and accurate in between
these two limiting cases. Since the calculations are more complicated than
before we refer the reader to our previous work for details \cite{Previos}.%
\begin{figure}
[ptb]
\begin{center}
\includegraphics[
height=9.3207cm,
width=8.4097cm
]%
{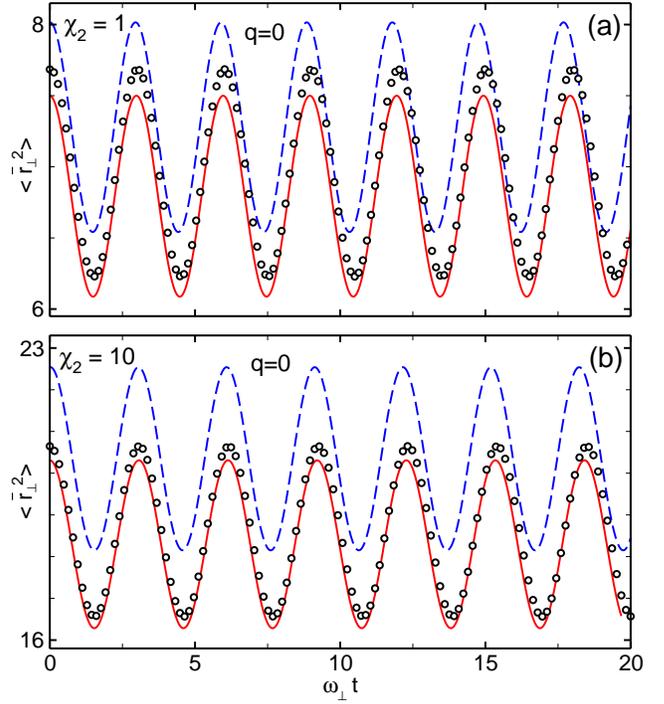}%
\caption{(Color online) Time evolution of the mean squared amplitude
$\langle\overline{r}_{\bot}^{2}\rangle$ of disk-shaped condensates with
$\chi_{2}=1$ and $10$ after a perturbation $\omega_{\bot}\rightarrow
1.1\omega_{\bot}$. The aspect ratio of the harmonic trap before perturbation
is $\lambda=\omega_{z}/\omega_{\bot}=10$. Solid lines are numerical results
obtained from our effective 2D equation (\ref{III-6}). Dashed lines are
numerical results obtained from the 2D NPSE by Salasnich \emph{et al.} Open
symbols are exact numerical results obtained from the full 3D GPE.}%
\label{Fig8}%
\end{center}
\end{figure}

Next we shall consider disk-shaped condensates in a confining potential that
is also harmonic in the radial direction $V_{\bot}(r_{\bot})=\frac{1}%
{2}m\omega_{\bot}^{2}r_{\bot}^{2}$. The corresponding oscillator length is
$a_{\bot}=\sqrt{\hbar/m\omega_{\bot}}$ and the trap aspect ratio now satisfies
the inequality $\lambda=\omega_{z}/\omega_{\bot}\gg1$. In Fig. \ref{Fig7} we
show the radial density profile $n_{2}(r_{\bot})$ of condensates in a trap
with $\lambda=10$, in the different relevant regimes characterized by the
parameter $\chi_{2}$. The dimensionless variable $\overline{r}_{\bot}$ is
defined as $\overline{r}_{\bot}=r_{\bot}/$ $a_{\bot}$. Figures \ref{Fig7}%
(a)--\ref{Fig7}(c) correspond to the ground-state ($q=0$) equilibrium
configuration, while Fig. \ref{Fig7}(d) corresponds to the equilibrium
configuration compatible with a $q=1$ vortex. Solid lines are numerical
results obtained from our effective 2D equation (\ref{III-6}) with $\mu
_{z}(|\varphi|^{2})$ given by Eqs. (\ref{III-15}) and (\ref{III-16}). Dashed
lines are numerical results obtained from the 2D NPSE by Salasnich \emph{et
al.}\ \cite{Reatto1} and open symbols are exact numerical results obtained
from the full 3D GPE. In the perturbative regime, for $\chi_{2}=0.1$, the 2D
NPSE is somewhat more accurate. This is precisely the parameter region where
the error of our effective equation is maximum. As $\chi_{2}$ increases, our
equation becomes more accurate. This can be appreciated in Figs.
\ref{Fig7}(b)--\ref{Fig7}(d) which show that our results remain very accurate
as $\chi_{2}$ increases. On the contrary, the error of the 2D NPSE increases
with $\chi_{2}$, being of the order of $5\%$ for $\chi_{2}=1$ and of the order
of $10\%$ for $\chi_{2}=10$. This is a consequence of the fact that the
underlying (Gaussian) variational wave function that leads to the 2D NPSE
cannot reproduce properly the TF regime.

To investigate the validity of our effective equation in time-dependent
problems we have followed the same procedure as before. We start from the
equilibrium configuration of condensates confined in a harmonic trap with
$\lambda=10$. We then perturb the system by changing instantaneously the
radial confinement frequency as $\omega_{\bot}\rightarrow1.1\omega_{\bot}$ and
follow the subsequent evolution in time of the mean squared radial amplitude
$\langle\overline{r}_{\bot}^{2}\rangle=\int d^{2}\overline{\mathbf{r}}_{\bot
}\,\overline{r}_{\bot}^{2}n_{2}(\overline{r}_{\bot},t)$. Figure \ref{Fig8}
shows the corresponding results for condensates with $\chi_{2}=1$ and $10$.
Solid lines are numerical results obtained from our effective 2D equation
(\ref{III-6}) with $\mu_{z}(|\varphi|^{2})$ given by Eqs. (\ref{III-15}) and
(\ref{III-16}), dashed lines are numerical results obtained from the 2D NPSE,
and open symbols are exact numerical results obtained from the full 3D GPE. As
is apparent from the figure, our effective 2D equation is again more accurate
than the 2D NPSE by Salasnich \emph{et al.}

\section{IV. CONCLUSION}

In this work, by applying the standard adiabatic approximation and using the
analytical expression for the transverse local chemical potential obtained in
our previous work \cite{Previos}, we have derived an effective 1D equation
that governs the axial dynamics of mean-field cigar-shaped condensates with
repulsive interatomic interactions. This equation, which incorporates
accurately the contribution from the transverse degrees of freedom, has the
correct limits in both the quasi-1D mean-field regime and the TF regime. Since
our expression for the local chemical potential is very accurate, the validity
of the above equation relies almost exclusively on the validity of the well
established adiabatic approximation.

We have compared our effective 1D equation with the 1D NPSE by Salasnich
\emph{et al.} which provides more accurate results than any other previously
proposed effective equation. We have demonstrated that our effective 1D
equation is more accurate, which, in part, is a consequence of the fact that
our approach, unlike that by Salasnich \emph{et al.}, reproduces correctly the
TF limit. Moreover, our equation allows treating condensates containing an
axisymmetric vortex with no additional cost. Because of its simplicity, it
also permits one to derive fully analytical expressions for ground-state
properties such as the chemical potential, axial length, axial density
profile, and local sound velocity. These analytical expressions reduce to the
correct analytical formulas in both the TF and the perturbative regimes, and
remain valid and accurate in between these two limiting cases.

Following the same procedure we have also derived an effective 2D equation
that governs the transverse dynamics of mean-field disk-shaped condensates,
accounting properly for the contribution from the axial degrees of freedom.
This equation is also more accurate than the 2D NPSE by Salasnich \emph{et al.
}As in the 1D case, from this effective equation, which also has the correct
limits in both the quasi-2D and the TF regime, one can derive analytical
expressions for ground-state properties of cigar-shaped condensates such as
the chemical potential, condensate radius, radial density profile, and local
sound velocity.

\begin{acknowledgments}
This work has been supported by MEC (Spain) and FEDER fund (EU) (Contract No. Fis2005-02886).
\end{acknowledgments}

\end{document}